\documentclass[journal,twoside]{IEEEtran}

\usepackage{setspace}
\usepackage[cmex10]{amsmath}
\usepackage{amssymb}  
\usepackage{graphicx} 
\usepackage{epstopdf}
\usepackage{bm}
\usepackage{balance}
\usepackage{lipsum}
\usepackage{array}  
\usepackage{color}
\usepackage{multirow}  
\usepackage{amsmath}
\usepackage{cite}
\usepackage{mathtools}
\usepackage{authblk}

\makeatletter
\def\blfootnote{\xdef\@thefnmark{}\@footnotetext}
\makeatother
\begin{document}
%\linenumbers

\title {\huge Reconfigurable Intelligent Surface-Assisted Space Shift Keying}
\author{Ayse E. Canbilen, Ertugrul Basar, ~\IEEEmembership{Senior Member,~IEEE}, Salama S. Ikki,~\IEEEmembership{Senior Member,~IEEE}

\thanks{A. E. Canbilen is with the Electrical and Electronics Engineering Department, Konya Technical University, Selcuklu 42250, Konya, Turkey (e-mail: aecanbilen@ktun.edu.tr).} %
\thanks{E. Basar is with Communications Research and Innovation Laboratory (CoreLab), Department of Electrical and Electronics Engineering, Ko\c{c} University, Sariyer 34450, Istanbul, Turkey  (e-mail: ebasar@ku.edu.tr).} %
\thanks{S. S. Ikki is with the Department of Electrical Engineering, Lakehead University, Thunder Bay, ON, P7B 5E1, Canada (e-mail: sikki@lakeheadu.ca).}
}

\maketitle

\begin{abstract}
The concept of reconfigurable intelligent surface (RIS)-assisted transmission, where phases of reflected signals are adjusted in an effective arrangement, has been recently put forward as a promising solution for 6G and beyond networks. Considering this and the undeniable potential of index modulation (IM) techniques, an RIS-based space shift keying (SSK) scheme is proposed in this letter to maintain all inherent advantages of both RISs and SSK. Specifically, a mathematical framework is presented by using a maximum likelihood (ML) detector for the calculation of the average bit error rate (ABER). Extensive computer simulation results are provided to assess the potential of the proposed RIS-based SSK system and to verify the theoretical derivations. The obtained results indicate that the proposed scheme  enables highly reliable transmission with unconventionally high energy efficiency, however with the added cost of increased receiver complexity.
 
 \end{abstract}
 \vspace{-.3cm}

\section{Introduction} 

There has been a quantum leap in the spectrum efficiency of wireless networks over the last few decades. It is foreseeable that the fifth generation (5G) standard, which will use frequency bands above 6 GHz, is going to provide more flexibility in system design, thanks to various technological advances such as massive multiple-input multiple-output (MIMO) and millimeter wave (mmWave) communications. Going one step further, the introduction of sixth generation (6G) mobile communication systems is expected around 2030 \cite{1}. 

Although future 6G technologies may simply seem to be the extension of their 5G counterparts, the immense jump to higher frequency bands, including the Terahertz (THz) band, requires an actual re-engineering of the whole system \cite{1}. New user requirements, completely new applications and epochal networking trends of 2030 and beyond may bring more challenging engineering problems. Hence, devices, transceivers and network-related technologies need to be designed and/or optimized for these very-high frequency bands. In addition to this, providing high performance wireless and mobile connectivity to all users, in an energy, spectrum and cost-effective manner is one of the key challenges of 6G \cite{1}. 

Since new technologies are approaching with rather rapid footsteps and users expect better and more versatile services, researchers relentlessly keep exploring the potential of recent advances such as index modulation (IM) and THz communications for 5G and beyond \cite{3,5}. Among these novel schemes, the emerging IM concept, which has been widely recognized by both academia and industry over the past few years, uses the indices of the available transmit entities to convey additional information bits in rich scattering environments \cite{7,8,9}. For instance, spatial modulation (SM), as an IM-based method, utilizes the index of an activated transmitter (Tx) antenna to convey extra information \cite{9}. This unique property allows SM to strike a favorable trade-off between spectral and energy efficiency. Moreover, it cancels out any inter-channel interference at the receiver (Rx), while simultaneously solving synchronization problems \cite{11}.

Researchers have also recently focused on controlling the propagation environment to increase the quality of service and/or spectrum efficiency, inspired by the idea of intelligent meta-surfaces \cite{12}. Regarding the concept of reconfigurable intelligent surface (RIS), artificial thin films of electromagnetic and reconfigurable materials are embedded into environmental objects/walls and controlled intentionally by a software to boost the signal quality at the Rx \cite{R2}. Indeed, an RIS only reflects the incident signal with an adjustable phase-shift, without requiring a dedicated energy source, decoding, encoding, or retransmission via a large number of small, low-cost, and passive elements on it \cite{15}. Although reflect-arrays do not buffer or process any incoming signals, the useful signals can be enhanced by optimally controlling the phase shift of each element on the reflect-arrays \cite{20}. Additionally, the RISs are affected neither by noise amplification nor by the effects of self-interference, as reflectors are not affected by such impairments \cite{R1}. 

From these aspects, the RIS concept substantially differs from existing MIMO and amplify-and-forward (AF) relaying paradigms. The potential of an RIS-assisted IM system to ensure high spectral efficiency (SE) at low SNR values was proven in \cite{18}, and it was shown in \cite{19} and \cite{chuang} that RISs provide significant gains in energy efficiency compared to conventional relay-assisted communications. An RIS-aided system can achieve the same data rate as a full-duplex/half-duplex AF relay-aided system by using more reflecting elements \cite{R222}. However, since the elements of RISs are passive, no RF chains are needed for them and thus the cost is much lower compared to that of active antennas for the AF relay requiring transmit RF chains \cite{R222}. All these advantages allow RIS-assisted communications to stand out as a promising field of research for beyond 5G. 

In this paper, the potential of IM-based transmission through an RIS has been investigated in terms of error probability and complexity while maintaining the inherent advantages of both IM and RIS. Specifically, an RIS-based space shift keying (SSK) scheme is proposed not only to provide an ultra-reliable and energy-efficient transmission, but also to avoid the synchronization and interference problems by exploiting the Tx antenna indices. Contrary to \cite{18}, the Tx antenna indices are exploited to realize SSK at the Tx side. An analytical framework, confirmed by extensive computer simulations, is provided by utilizing maximum likelihood (ML) detection and calculating the average bit error rate (ABER) of the proposed system. An asymptotic approximation is also derived to gain useful insights. Additionally, the effects of the number of reflecting elements, blind phases, channel phase estimation error and path loss on the overall system performance are investigated depending on different signal-to-noise ratio (SNR) regimes. 

\vspace{-.2cm}

\section{System Model} 

An RIS-based SSK transmission scheme, where the RIS is effectively used as a reflector in a dual-hop (DH) communication scenario with $N_t$ Tx antennas at the source (S) and one Rx antenna at the destination (D), is considered in this work and illustrated in Fig. 1. In this regard, SSK is realized at the Tx side by mapping the incoming bits into the index of a specific Tx antenna, which is indicated by $i$. This antenna is activated for transmission and it conveys the unmodulated carrier signal generated by S to the RIS to be reflected to D.
Assuming that the RIS has $N$ simple and reconfigurable reflector elements, $h_{l,i}$ and $g_{l}$ represent the channel fading coefficients between $i^{th}$ Tx antenna of S and the reflectors of the RIS, and the reflectors of the RIS and the receive antenna of D, respectively, where $l=1,...,N$ and $i\in \{1,...,N_t\}$. Additionally, $h_{l,i}$ and $g_{l}$ follow zero-mean complex Gaussian distribution with variance $\sigma^2$. 

The general ray-tracing (GRT) model in \cite{gold} can be utilized to assess the path loss effect on the proposed RIS-based SSK system design. Accordingly, it can be assumed that all the scattered and diffracted signals are included in the small-scale fading model, and the reflections are caused by an RIS laid on the ground. Assume that the distance, $d$, between S and D is far greater than the sum of the ground clearance of S and D, the travelled distance of the reflected signals as well as the direct path component can be approximated to the distance between S and D \cite{gold}\footnote{$d \approx l_d \approx d_1+d_2$, where $d_1$ and $d_2$ represent the distance between the Tx and the center of the RIS, and the distance between the Rx and the center of the RIS, respectively, while $l_d$ is the length of the direct path.}. Considering all these assumptions with unit gain antennas, and noting that an unmodulated carrier signal is being conveyed through just one specific Tx antenna with SSK, the received power, $P_r$, can be written in terms of the transmitted power, $P_t$, as $P_r \approx P_t (N+1)^2(\lambda/4\pi d)^2$, where $\lambda$ being the wavelength. Ignoring the direct path propagation yields a received power of $P_r \approx P_t N^2(\lambda/4\pi d)^2$. Here, it can be concluded that the direct path would not have a major effect on the performance of the proposed system for large $N$, i.e., $N \gg 1$. Hence, it has not been considered in this study. On the other hand, when taking the path loss effects into account, $\sigma^2$ would depend on its value. Here, it is worth noting that an RIS can be viewed as a specular reflector if it is large enough (for example if both its length and width are 10 times larger than the wavelength) according to geometrical optics \cite{R2}. The experimental measurements validated that the free-space path loss of RIS-assisted wireless communication is proportional to $(d_1+d_2)^2$, as given here, under the scenario of specular reflection in the near field\footnote{When the distance between the Tx/Rx and the center of the RIS is less than $\frac{2A}{\lambda}$, the Tx/Rx is considered to be in the near field of the RIS, where $A$ denotes the total area of the RIS \cite{1881}.} broadcasting case. Otherwise, it is proportional to $(d_1d_2)^2$ \cite{1881}. 

The concept of RIS-based SSK is discussed in the following sections by considering two different implementation scenarios based on the knowledge of the channel phases at the RIS: intelligent RIS-SSK and blind RIS-SSK.  \vspace{-.1cm}

\subsection{Intelligent RIS-SSK}
In this scenario, the RIS is assumed to be controlled by a communication software to specify the reflection phases in such a way that the SNR at D is maximized. For this case, the received baseband signal at D is written as: \vspace{-.1cm}
    \begin{align}
    y=\sqrt{E}\bigg( \sum_{l=1}^{N} h_{l,i}e^{j \phi_l}g_{l} \bigg)+n, \label{eq1}  \vspace{-.3cm}
     \end{align}
where $E$ is the transmitted signal energy, $n$ is the zero-mean additive white Gaussian noise term with variance $N_0$, $ \phi_l$ is the adjusted phase for the $l^{th}$ reflector of the RIS and $i$ is the selected Tx antenna index. By defining $h_{l,i}=\alpha_{l,i}e^{-j\theta_{l,i}}$ and $g_{l}=\beta_{l}e^{-j\psi_{l}}$, the instantaneous SNR at D can be calculated from: \vspace{-.1cm}
  \begin{align}
    SNR=\frac{E | \sum_{l=1}^{N} \alpha_{l,i} \beta_{l}e^{j(\phi_l-\theta_{l,i}-\psi_l)} |^2}{N_0}. \label{eq2} 
     \end{align}
Here, it is easy to show that this SNR value can be maximized by eliminating the channel phases\footnote{This can be verified by the identity $|\sum_{a=1}^{N}z_ae^{j\xi_a}|^2=\sum_{a=1}^{N}z_a^2+2\sum_{a=1}^{N}\sum_{b=a+1}^{N}z_az_b\cos(\xi_a-\xi)$, which rises to maximum when $\xi_a=\xi$ for all $a$ \cite{18}.}. In other words, $\phi_l$ is adjusted as $\phi_l=\theta_{l,i}+\psi_l$ with the help of the RIS that has the knowledge of channel phases $\theta_{l,i}$ and $\psi_l$. Please note that this models the best-case scenario, in which the RIS also acquires the information of the activated Tx antenna index, which might be possible for the case where S is in the near field of the RIS. It is assumed that the channel state information is provided for the RIS via a communication software. Considering this, an ML detector, which determines the index of the activated Tx antenna, is designed for the intelligent RIS-SSK scheme as: \vspace{-.1cm}
  \begin{align}
    q=\arg \min_{i} \bigg\{  \bigg| y - \sqrt{E} \bigg( \sum_{l=1}^{N} \alpha_{l,i}\beta_{l} \bigg) \bigg|^2 \bigg\}. \label{eq3}
     \end{align} \vspace{-.3cm}

It is worth noting that more than one phase should be adjusted for each reflector of the RIS to cancel the channel phases in the case of multiple antennas at the Rx side, as considered in \cite{18}. However, it would not be a practical scenario for this design since the reflector elements can be set at just one specific phase for each reflection. \vspace{-.4cm}

\begin{figure}[t] \vspace{-1.5cm}
\centering
\includegraphics[height=6cm, width=8cm]{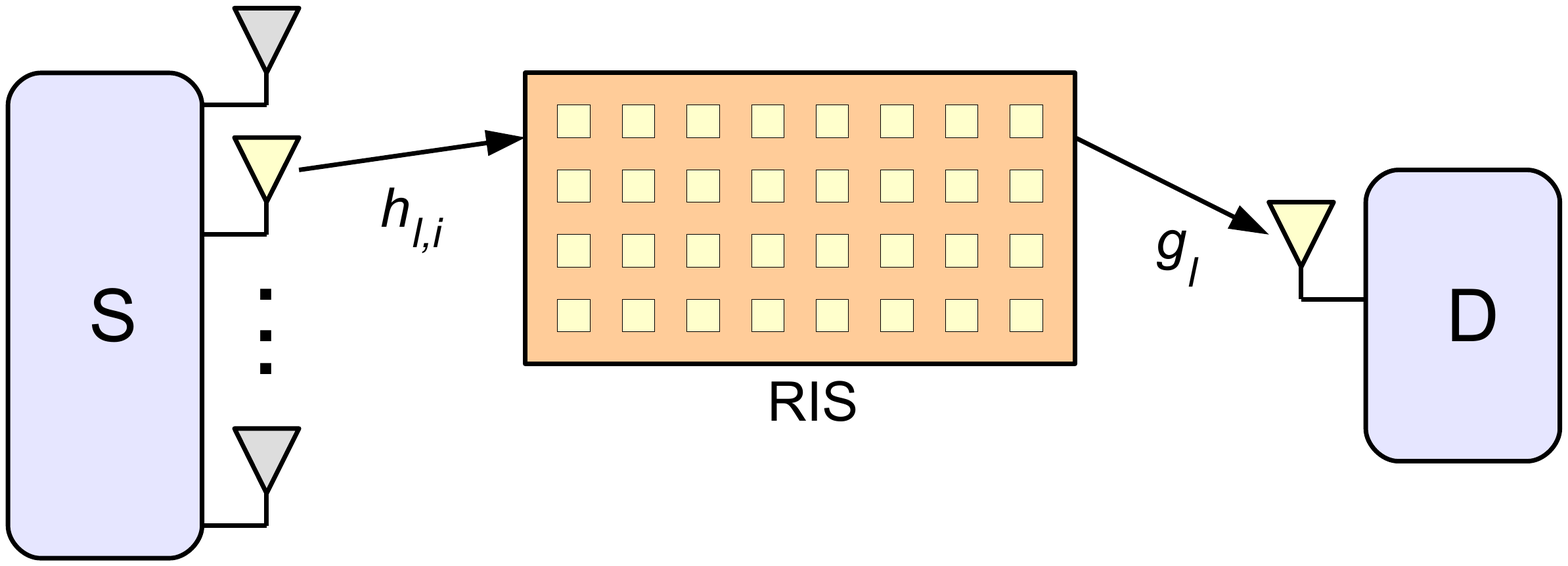} \vspace{-2.1cm}
\caption {RIS-assisted SSK system model.} \vspace{-.5cm}
\end{figure} 

\subsection{Blind RIS-SSK}
The strategy of maximizing the SNR value by adjusting phases for the reflectors cannot be implemented in this case, because the RIS has neither channel nor activated Tx antenna index information. Compared to the intelligent case, this models the worst-case scenario in terms of system performance. The RIS is exploited just for the blind reflection of the unmodulated carrier signal assuming that $\phi_l=0^\circ$ in \eqref{eq1} for all $l$. %, the received signal at D can be obtained as follows:
%  \begin{align}
 %   y=\sqrt{E}\bigg( \sum_{l=1}^{N} h_{l,i}g_{l} \bigg)+n. \label{eq4}
  %   \end{align}
For this blind scheme, an ML detector performs the detection of the activated Tx antenna index as follows: \vspace{-.1cm}
  \begin{align}
    q=\arg \min_{i} \bigg\{ \bigg| y - \sqrt{E}\bigg( \sum_{l=1}^{N} h_{l,i}g_{l} \bigg) \bigg|^2 \bigg\}. \label{eq5}
     \end{align} \vspace{-.2cm}

Note that multiple antennas can be considered at the Rx for this case, since $\phi_l$ does not depend on $l$. \vspace{-.2cm}
 
\section{Performance Analysis}
In this section, analytical derivations are presented for the calculation of the theoretical ABER of the proposed RIS-based SSK schemes over Rayleigh fading channels for both intelligent and blind transmission cases. \vspace{-.4cm}

\subsection{Performance of the Intelligent RIS-SSK} 
Assuming the selection of the Tx antenna $i$ and its erroneous detection as $q$, the conditional pairwise error probability (PEP) of the intelligent RIS-SSK is obtained by using the decision rule in \eqref{eq3} as follows: \vspace{-.1cm}
  \begin{align} 
P_e\hspace{-.1cm}=\hspace{-.1cm}Pr\left\{i \rightarrow q |G\right\} \hspace{-.1cm}=Pr\left\{ | y \hspace{-.1cm}-\hspace{-.1cm} \sqrt{E}G_i |^2\hspace{-.1cm}>\hspace{-.1cm} | y\hspace{-.1cm} -\hspace{-.1cm} \sqrt{E}{G}_q |^2 \right\}, \hspace{-.1cm} \vspace{-.2cm} \label{eq6} 
     \end{align} 
where $G_i=\sum_{l=1}^{N} \alpha_{l,i}\beta_{l}$, $G_q=\sum_{l=1}^{N} \alpha_{l,q}\beta_{l}e^{-j(\theta_{l,q}-\theta_{l,i})}$, and $Pr \{ \cdot \}$ denotes the probability of the event. 

After some mathematical operations, \eqref{eq6} can be written as $P_e=Pr\{ D>0 \}$, where $D=-E|G_i-G_q|^2-2 \Re \{ \sqrt{E}(G_i-G_q) n^*\}$ is a complex Gaussian random variable (RV) with mean $\mu_D$ and variance $\sigma_D^2$, while $n^*$ is the complex conjugate value of $n$. Noting that $\mu_D=-E |G_i-G_q|^2$ and $\sigma_D^2=2EN_0|G_i-G_q|^2 $, the conditional PEP can also be calculated by using the well-known $Q$-function: 
  \begin{align}
    P_e=Q\left( \sqrt{\frac{E |G_i-G_q|^2}{2N_0}} \right) . \label{eq8}
     \end{align}
Here, by denoting $B=G_i-G_q$, it can be expanded as follows: \vspace{-.2cm}
 \begin{align}
B=\sum_{l=1}^{N} \beta_{l} (\alpha_{l,i}-\alpha_{l,q}e^{-j(\theta_{l,q}-\theta_{l,i})}). \vspace{-.2cm}\label{eq9}
  \end{align}
By defining $\beta=\beta_{l}$ and $\alpha=\alpha_{l,i}-\alpha_{l,q}e^{-j(\theta_{l,q}-\theta_{l,i})}$ in \eqref{eq9}, it can be observed that $\beta$ and $\alpha$ are independent RVs. At this point, the Central Limit Theorem (CLT) can be advantageously used for a sufficiently large number of reflecting elements, $N\gg1$. According to the CLT, $B$ follows Gaussian distribution with the following mean and variance values regardless of the distributions of its components\footnote{If $X$ and $Y$ are independent RVs, $\mathrm{Var}\{XY\}=\mathrm{Var}\{X\} \mathrm{Var}\{Y\}+\mathrm{Var}\{X\} \mathrm{E}^2\{Y\}+\mathrm{E}^2\{X\} \mathrm{Var}\{Y\}$ and $\mathrm{E}\{XY\}=\mathrm{E}\{X\} \mathrm{E}\{Y\}$. Beside, if $R$ follows Rayleigh distribution, $\mathrm{E}\{R\}= \sqrt{\sigma^2\pi/2}$ and $\mathrm{Var}\{R\}=\sigma^2(4-\pi)/2$, where $\sigma^2$ denotes the variance of the Gaussian RVs that generate $R$.}:
 \begin{align}
\mu_B=\frac{N \pi}{4}, \;\;\;\;\; \sigma_B^2=\frac{N(32-\pi^2)}{16}.
 \end{align}

In order to find the average PEP (APEP), the following should be calculated:
 \begin{align}
\overline{P}_e=\int_0^\infty Q \bigg( \sqrt{ \frac{E \lambda}{2 N_0} }\bigg) f_{\lambda}(\lambda)d\lambda , \label{eq11}
  \end{align}
where $\lambda= |B|^2$. Utilizing the moment-generating function (MGF) of $\lambda$, which follows the non-central chi-square distribution, the APEP is obtained as:
  \begin{align}
\overline{P}_e= \frac{1}{\pi} \int_0^{\pi/2} M_{\lambda}\bigg( -\frac{E }{4N_0 \sin^2\eta}\bigg)d\eta, \label{eq12}
     \end{align}
where $M_\lambda(\cdot)$ is the MGF of $\lambda$, which is defined as follows:
  \begin{align}
M_\lambda (t)= \bigg(\frac{1}{\sqrt{1-\sigma_B^2t}}\bigg)\exp\bigg(\frac{\mu_B^2t}{1-\sigma_B^2t}\bigg) . \vspace{-.2cm} \label{eq13}
     \end{align}

Considering that the maximum value of $\sin \eta$ is 1, \eqref{eq12} can be upper-bounded to gain insights into the behavior of the intelligent RIS-SSK scheme in terms of error probability as: \vspace{-.1cm}
\begin{align}
  \hspace{-.2cm}\overline{P}_e\approx \frac{1}{2}\bigg(\hspace{-.1cm}1+\frac{NE(32-\pi^2)}{64N_0} \bigg)^{-\frac{1}{2}}\hspace{-.1cm}\exp \bigg( \frac{\frac{-N^2\pi^2E}{64N_0}}{1+\frac{NE(32-\pi^2)}{64N_0}} \bigg). \hspace{-.1cm}\label{eq14}
     \end{align}
This expression reveals interesting as well as important points given in the following remarks.

{\it Remark} 1: Assuming that $\frac{NE}{N_0}\ll 3$ in \eqref{eq14}, the APEP value becomes proportional to:  \vspace{-.1cm}
\begin{align}
\overline{P}_e \propto \exp \bigg( \frac{-N^2\pi^2E}{64N_0} \bigg). \label{eq15}
\end{align}

It is clear from \eqref{eq15} that incredibly low APEP values can be achieved by increasing $N$, owing to the $N^2$ term in the exponential form, even if the SNR $(E/N_0)$ is relatively low.

{\it Remark} 2: Considering the high SNR values, an asymptotic approximation can be derived by using \eqref{eq14} for the APEP as:  \vspace{-.4cm}
\begin{align}
  \hspace{-.2cm}\overline{P}_e\approx \bigg(\hspace{-.1cm}\frac{NE(32-\pi^2)}{16N_0} \bigg)^{-\frac{1}{2}}\hspace{-.1cm}\exp \bigg( \frac{-N\pi^2}{32-\pi^2} \bigg). \hspace{-.1cm}\label{eq16}
     \end{align}

It can be inferred from \eqref{eq16} that the error probability reduces exponentially with increasing $N$ for high SNR values. On the other side, increasing the SNR also decreases the APEP value; however, this reduction is relatively small due to the exponent being $-\frac{1}{2}$ .  \vspace{-.2cm}

\subsection{Performance of the Blind RIS-SSK}

The ML detector performs the detection of the activated antenna by using the detection rule given in \eqref{eq5} for the blind RIS-SSK. 
Assuming that the Tx antenna $i$ has been activated for transmission, although it is detected as $q$ at D, the PEP is calculated from:  \vspace{-.1cm}
  \begin{align}
P_e=Pr\left\{| y - \sqrt{E}K_i |^2>| y - \sqrt{E}K_q |^2 \right\}, \label{eq17}
     \end{align}
where $K_i=\sum_{l=1}^{N} h_{l,i}g_{l}$ and $K_q=\sum_{l=1}^{N} h_{l,q}g_{l}$. Following a similar analysis, the conditional PEP can be written as:  \vspace{-.1cm}
  \begin{align}
    P_e=Q\left( \sqrt{\frac{E|K_i-K_q|^2}{2N_0}} \right) . \label{eq18}
     \end{align}
Here, defining $\delta=g_{l}$ and $\Psi=h_{l,i}-h_{l,q}$, which are clearly independent RVs, $K_i-K_q=C=\sum_{l=1}^{N} \delta \Psi$. 
According to CLT, $C$ follows a zero-mean Gaussian distribution with the variance of $\sigma_C^2=2N$.  
 
Noting that $|K_i-K_q|^2$ in \eqref{eq18}, which is equal to $|C|^2$, follows the central chi-square distribution, the APEP can be obtained for the blind RIS-SSK as follows:
  \begin{align}
  \overline{P}_e= \frac{1}{2} \left(1-\sqrt{\frac{2N(E/2N_0)}{2+2N(E/2N_0)}} \; \right). \label{eq21} 
     \end{align}  \vspace{-.6cm}

\subsection{Average Bit Error Rate (ABER)}
It is important to mention that the ABER is equal to the APEP for binary signaling, i.e., while $N_t=2$. For the general case of $N_t>2$, the following union bound can be used to calculate the ABER value \cite{18}:  \vspace{-.1cm}
  \begin{align}
 \overline{P}_b \leq \frac{N_t}{2} \overline{P}_e, \label{eq22}  
     \end{align}
which considers the fact that the APEP is independent of activated antenna indices, and identical for all pairs.  \vspace{-.2cm}

\section{Detection Complexity Analysis}
In order to assess the computational complexity, real multiplications and summations required for the detection procedure are counted. Considering \eqref{eq3}, the complexity of intelligent RIS-SSK scheme is obtained as $N+3$ real multiplications and $N+1$ real summations. On the other hand, \eqref{eq5} is utilized to obtain the complexity of the blind RIS-SSK scheme with similar calculations, and it is concluded that $4N+4$ real multiplications and $3N+2$ real summations are needed in order to detect the activated transmit antenna. 

The results of the complexity analysis are summarized in Table 1, where $M$ and $S$ denote the real multiplications and summations, respectively. The complexity of the conventional SSK \cite{sal} is included for comparison as well. Keeping the assumption of $N\gg1$, it is evident from this table that the conventional SSK is the simplest method, while the blind RIS-SSK is the most complex one. Moreover, the process of adjusting the phases of the reflecting elements at the RIS provides a significant gain in complexity compared to the blind scheme, especially for a rather large number of reflectors.  \vspace{-.2cm}

\begin{table}[t] \footnotesize 
\caption{Complexity Analysis Summary} \vspace{-.3cm}
\begin{center}
\begin{tabular}{ |c|c| } 
\hline
 \textbf{Scheme} &  \textbf{Complexity} \\
\hline
Conventional SSK &  $4S+4M$ \\
\hline
Blind RIS-SSK  & $(3N+2)S+(4N+4)M$  \\ 
\hline
Intelligent RIS-SSK & $(N+1)S+(N+3)M$  \\
\hline
\end{tabular}
\end{center} \vspace{-.6cm}
\end{table}

\section{Simulation Results}
In this section, extensive computer simulation results are provided to investigate the error performance of the proposed RIS-based SSK scheme in the context of both intelligent and blind transmission cases. The simulation results are verified by analytical results, and compared to reference schemes. The SNR is defined as $E/N_0$, and any path loss effect is neglected unless otherwise stated with $\sigma^2=1$. 

The error performance of the intelligent RIS-SSK scheme is presented for different numbers of reflecting elements in Fig. 2, where an AF-aided SSK scheme that includes a conventional $N$-antenna AF relay instead of an RIS structure is also considered. As seen from this figure, the proposed system achieves very low ABER at extremely low SNR values with increasing $N$. It is confirmed by Fig. 2 that the CLT approximation, which is utilized during the derivation of the theoretical APEP, is considerably accurate since the analytical and simulation results match perfectly for $N\geq32$. Clearly, the performance enhancement of the intellgent RIS-SSK scheme is far greater than that achieved by using the reference AF-SSK scheme with increasing $N$. It is observed that doubling the number of reflectors ensures 6 dB (four-fold) gain in the SNR while achieving a target ABER value at the waterfall region of the RIS-SSK, and this can be easily substantiated from \eqref{eq15}.

In Fig. 3, the ABER performance of the intelligent RIS-SSK scheme is compared to the traditional SSK, in which the data is  transmitted from S to D without an RIS. It is observed that the intelligent scheme provides considerably better results. For instance, the proposed system with $N=16$ reflecting elements has approximately 30 dB (thousand-fold), and even more with increasing $N$, SNR gain for ABER$=10^{-3}$. The exact ABER is in compliance with its upper-bound, which is derived in \eqref{eq14}. On the other hand, the ABER curves of the intelligent RIS-SSK have a waterfall region as well as a saturation region. At this point, note that this figure confirms both Remarks 1 and 2, which explain the superior ABER performance of the RIS-based SSK especially with increasing $N$, and the saturated ABER performance for high SNR values, respectively. In Fig. 3, the intelligent RIS-based SSK scheme is also compared to the intelligent RIS-based DH communication system proposed in \cite{15} in terms of error probability. It is known that the SE of DH transmission is calculated by $\log_2{M}$ for $M$-ary modulation and it is equal to $\log_2{N_t}$ for the SSK design. Considering this, it is observed that increasing the SE from 2 bps/Hz to 6 bps/Hz costs approximately 12 dB loss in SNR for the RIS-DH design, while it costs only 2 dB for the RIS-SSK. Hence, it can be concluded that the intelligent RIS-SSK method is considerably more energy-efficient for the systems that require high SE, such as 6G technologies, with the cost of increased number of Tx antennas.

\begin{figure}[t]  \vspace{-3.8cm} \hspace{-1.4cm}
\includegraphics[height=11cm, width=7.1cm]{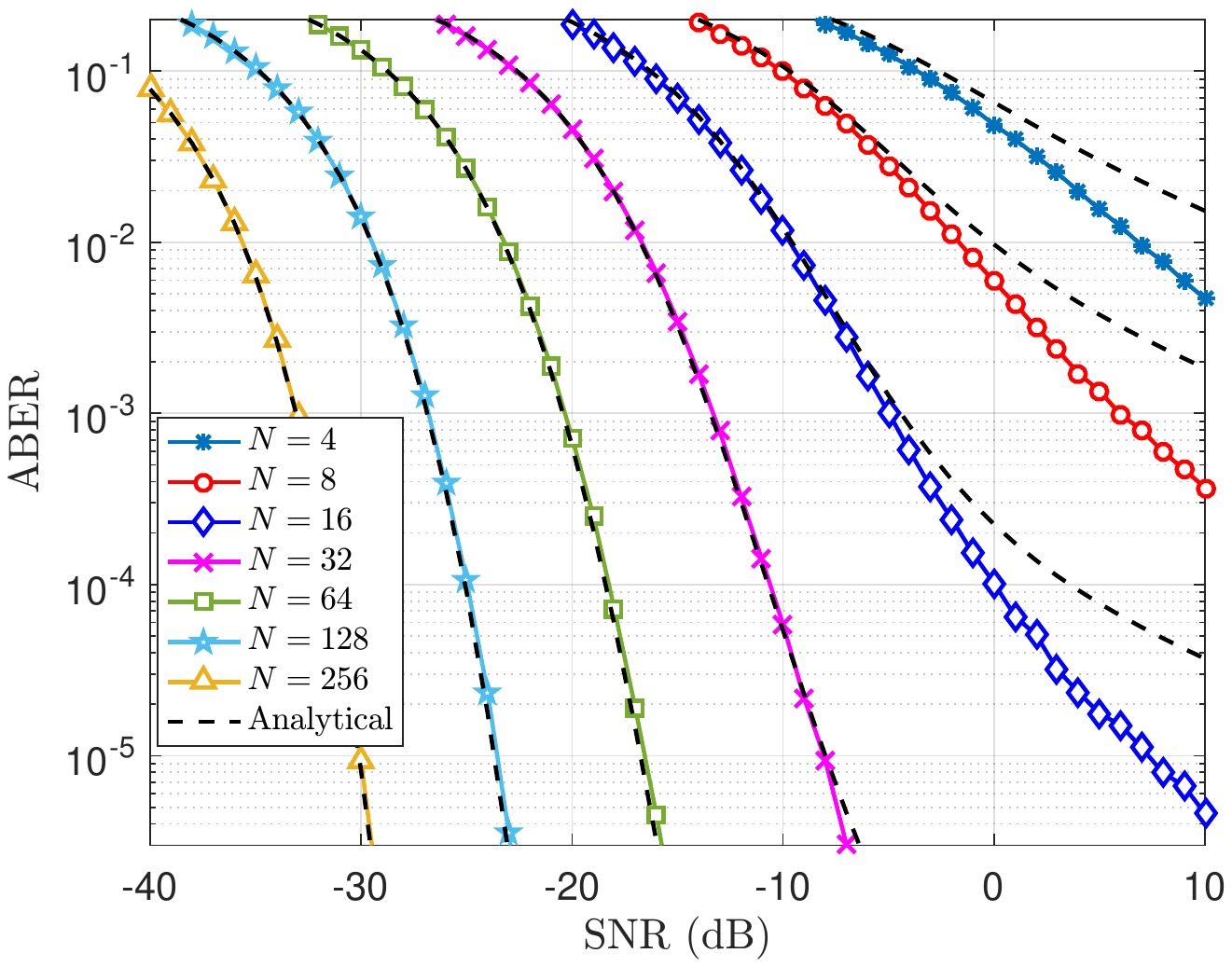} \hfill  \hspace{-5cm}
\includegraphics[height=11cm, width=7.1cm]{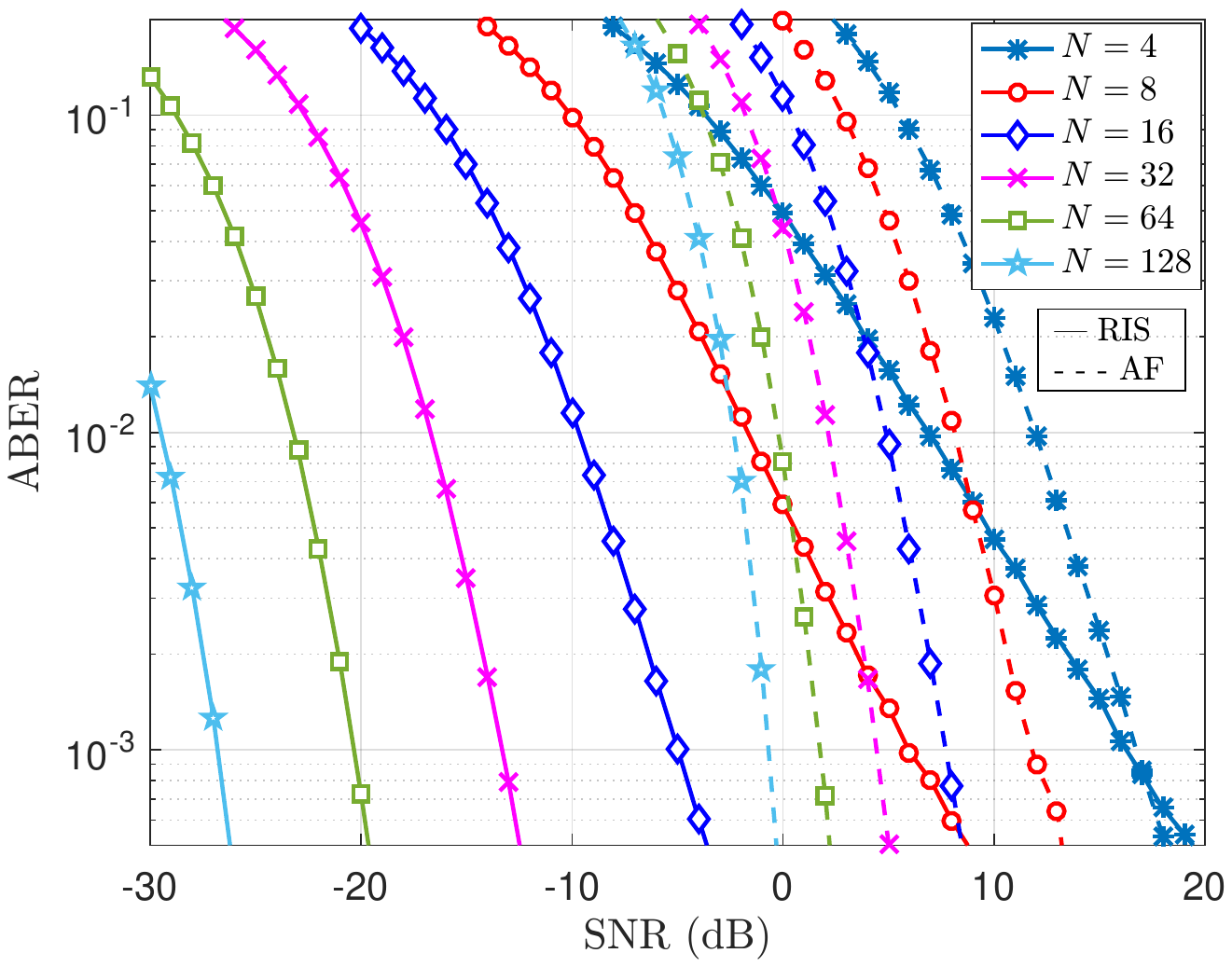} \hspace{-1.8cm} \vspace{-3.8cm}
\caption{ABER performance of the intelligent RIS-SSK scheme with increasing $N$ (left) and comparison with AF-aided SSK scheme ($N_t = 2$) (right).} \vspace{-.6cm}
\end{figure}

\begin{figure}[t]  \vspace{-3.3cm} \hspace{-1.4cm}
\includegraphics[height=11cm, width=7.1cm]{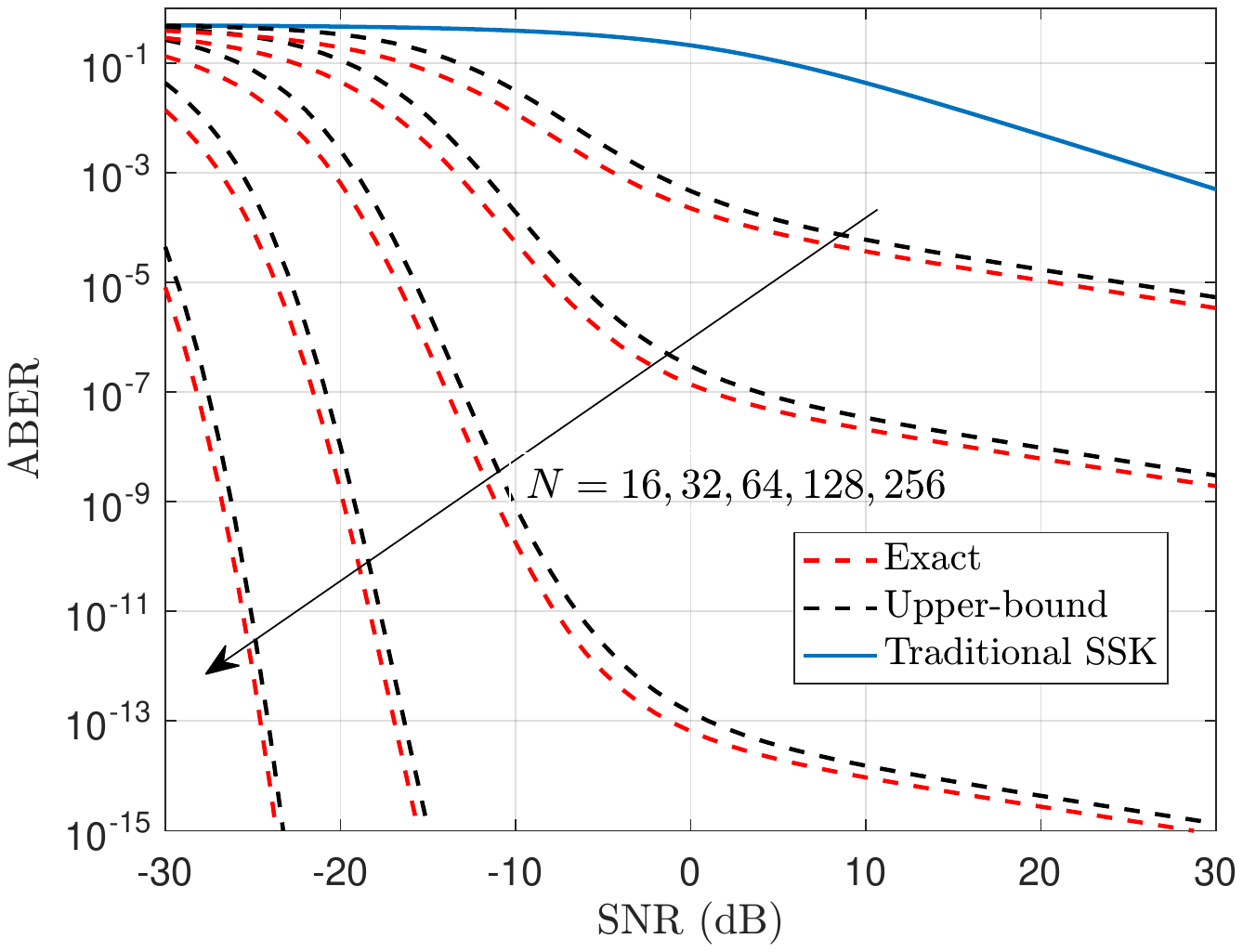} \hfill  \hspace{-5cm}
\includegraphics[height=11cm, width=7.1cm]{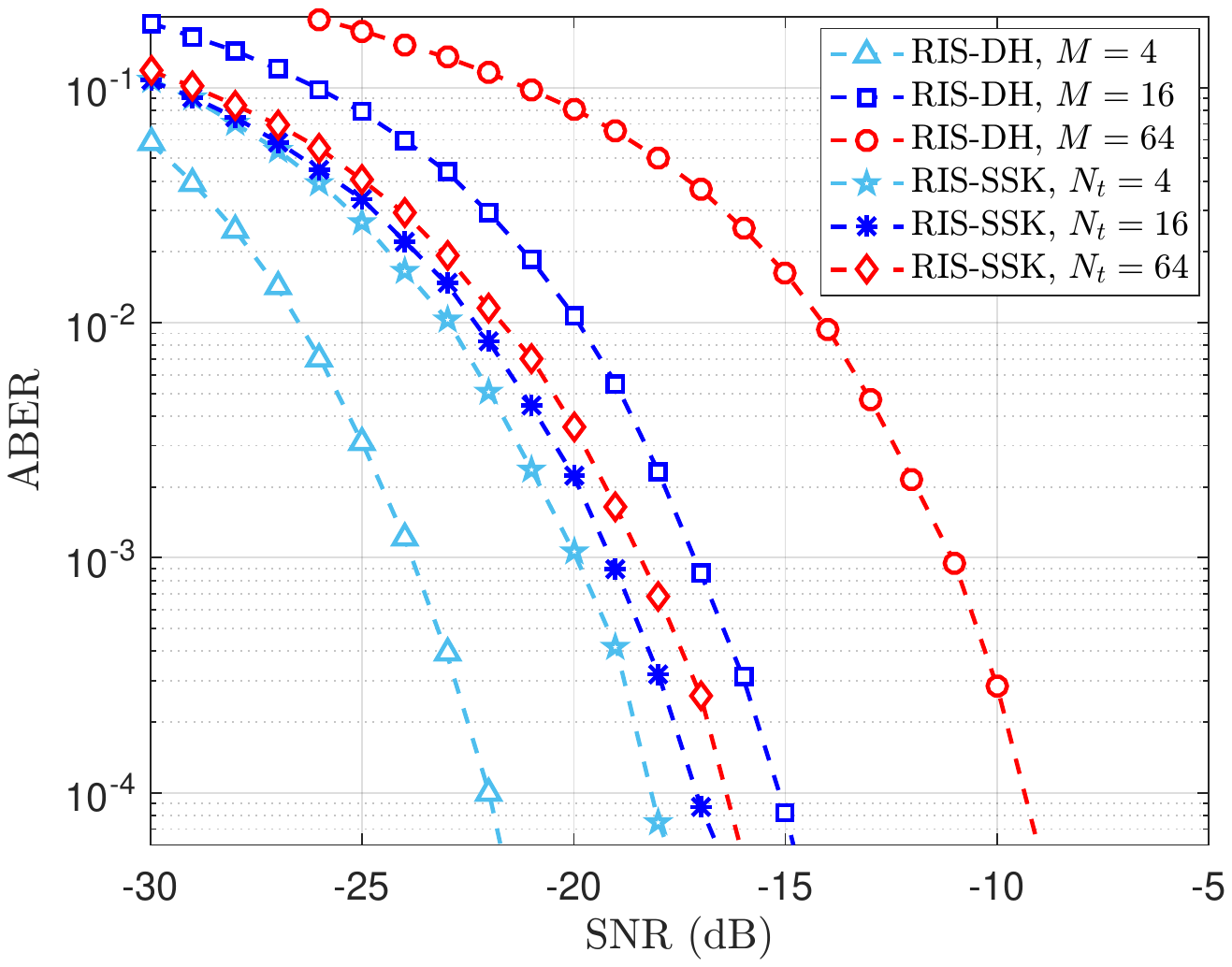} \hspace{-1.8cm} \vspace{-3.8cm}
\caption{ABER performance comparison of the intelligent RIS-SSK with traditional SSK for $N_t = 2$ (left), and the intelligent RIS-DH scheme proposed in [10] for $N = 64$ (right).} \vspace{-.5cm}
\end{figure}

Fig. 4 is provided to give insights about the performance of the blind RIS-SSK. Analytical ABER curves are also included in the same figure to check the accuracy of the theoretical findings. It is shown that the RIS-SSK exhibits $12$ dB better results than the traditional SSK when $N=16$, and even better for higher numbers of reflecting elements. Although the performance enhancement of the blind RIS scheme is parallel to the traditional method with an incresing number of Rx antennas, it provides 15 dB SNR reduction. Furthermore, considering both Figs. 2 and 4, it can be concluded that making the RIS-SSK intelligent by adjusting reflector phases enables a target ABER at incredibly lower SNR values.

The path loss effect on the performance of the traditional SSK is compared to the intelligent RIS-assisted SSK system in Fig. 5. For the proposed intelligent scheme, $\sigma^2$ is defined as $N^2(\lambda/4\pi d)^2$; however, for the traditional SSK, $\sigma^2=(\lambda/4\pi d)^2$, where $\lambda$ is the wavelength and $d$ is the total distance. The given results are obtained for 28 GHz, which is one of the well-studied mmWave bands in the literature \cite{R4*} and provides $\lambda \approx1$ cm. Assuming an element spacing of $\lambda/2$, the total area of the RIS is defined as $A=N(\frac{\lambda}{2})^2$. Accordingly, the far field boundary  is calculated as 0.64 m, 1.28 m and 2.56 m for the $N$ values of 128, 256 and 512, respectively. Then, the RIS is assumed to be located at $d_1=0.6$ m under the scenario of specular reflection in the near field. It is observed that doubling $N$ provides 6 dB better results for the proposed scheme, which is definitely more robust to the path loss effects than the traditional SSK. The ABER performances are also compared with increasing S-D distances from 10 to 30 meters in Fig. 5. Obviously, the path loss has serious destructive effects on both systems. However, increasing the number of reflecting elements of the RIS remarkably enhances the system performance, and this proves the potential of RISs for beyond 5G technologies.

Lastly, the performance of the intelligent RIS-SSK scheme in the presence of channel phase estimation errors is presented in Fig. 6. The phase error at the RIS is modelled as a zero-mean von Mises variable, whose concentration parameter $\kappa$ captures the accuracy of the estimation \cite{18}. As illustrated in Fig. 6, even an imperfect knowledge of the ideal phase shifts can be useful. For instance,  when $\kappa=2$ (for which the distribution is broad)  the gap to perfect channel phase information is of about 3 dB, while for $\kappa=5$ (more accurate estimation) this value reduces to 1 dB. \vspace{-.3cm}

\begin{figure}[t]  \vspace{-3.8cm} \hspace{-1.4cm}
\includegraphics[height=11cm, width=7cm]{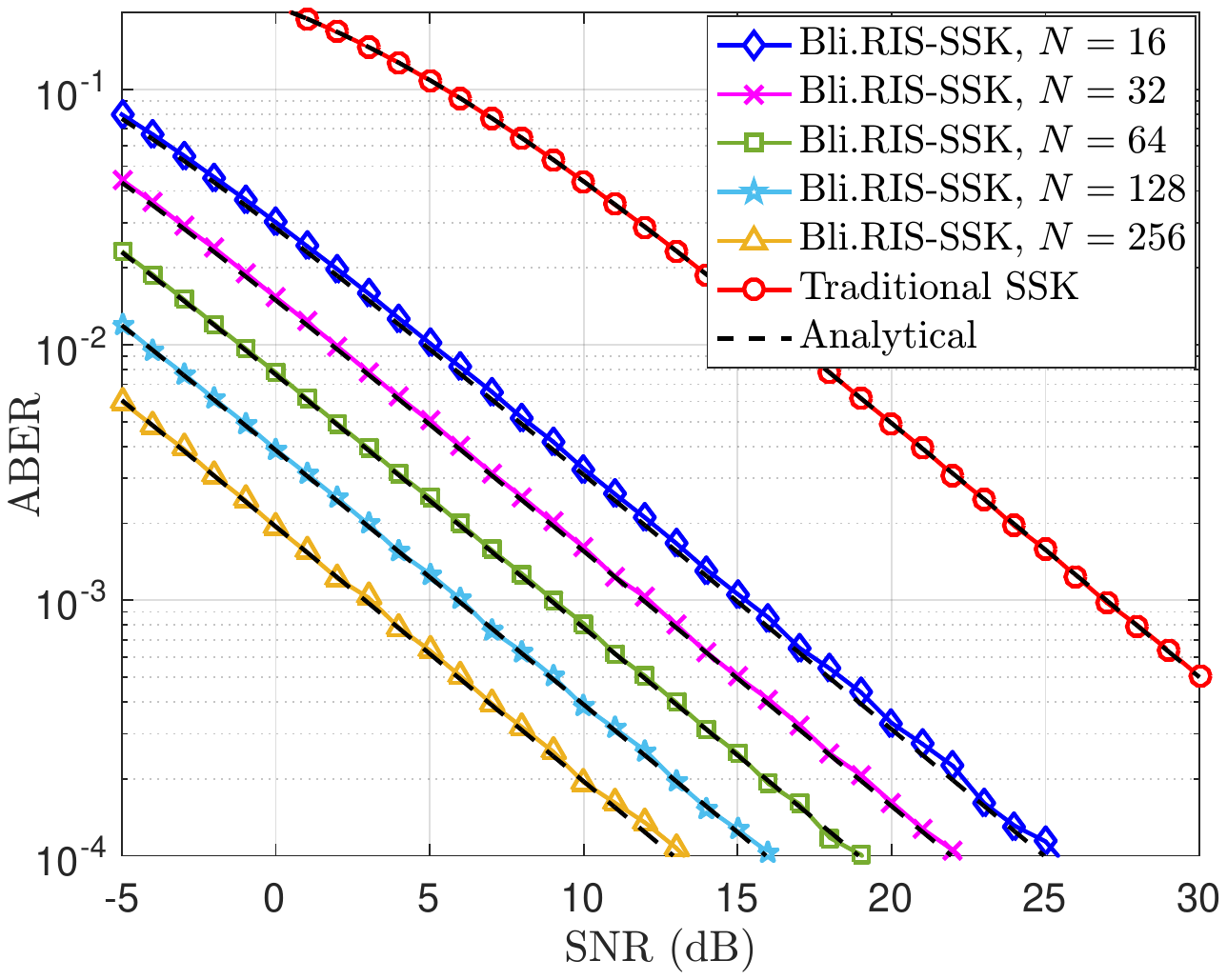} \hfill  \hspace{-5cm}
\includegraphics[height=11cm, width=7cm]{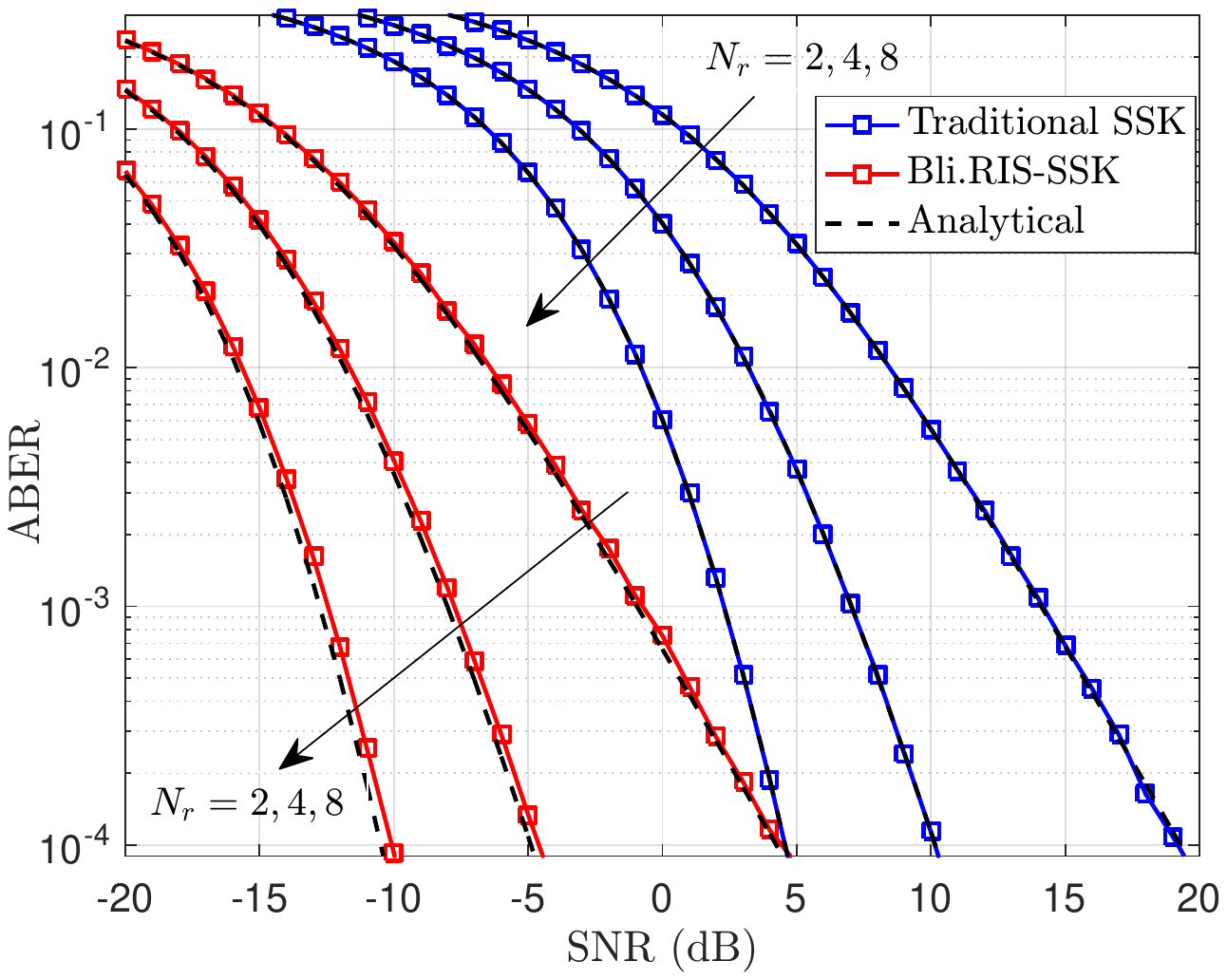} \hspace{-1.6cm} \vspace{-3.8cm}
\caption{ABER performance of the blind RIS-SSK and traditional SSK schemes ($N_t=2$) for varying $N$ with $N_r=1$ (left), and varying $N_r$ with $N=32$ (right).} \vspace{-.4cm}
\end{figure}

\begin{figure}[t]  \vspace{-3.5cm} \hspace{-1.4cm}
\includegraphics[height=11cm, width=7cm]{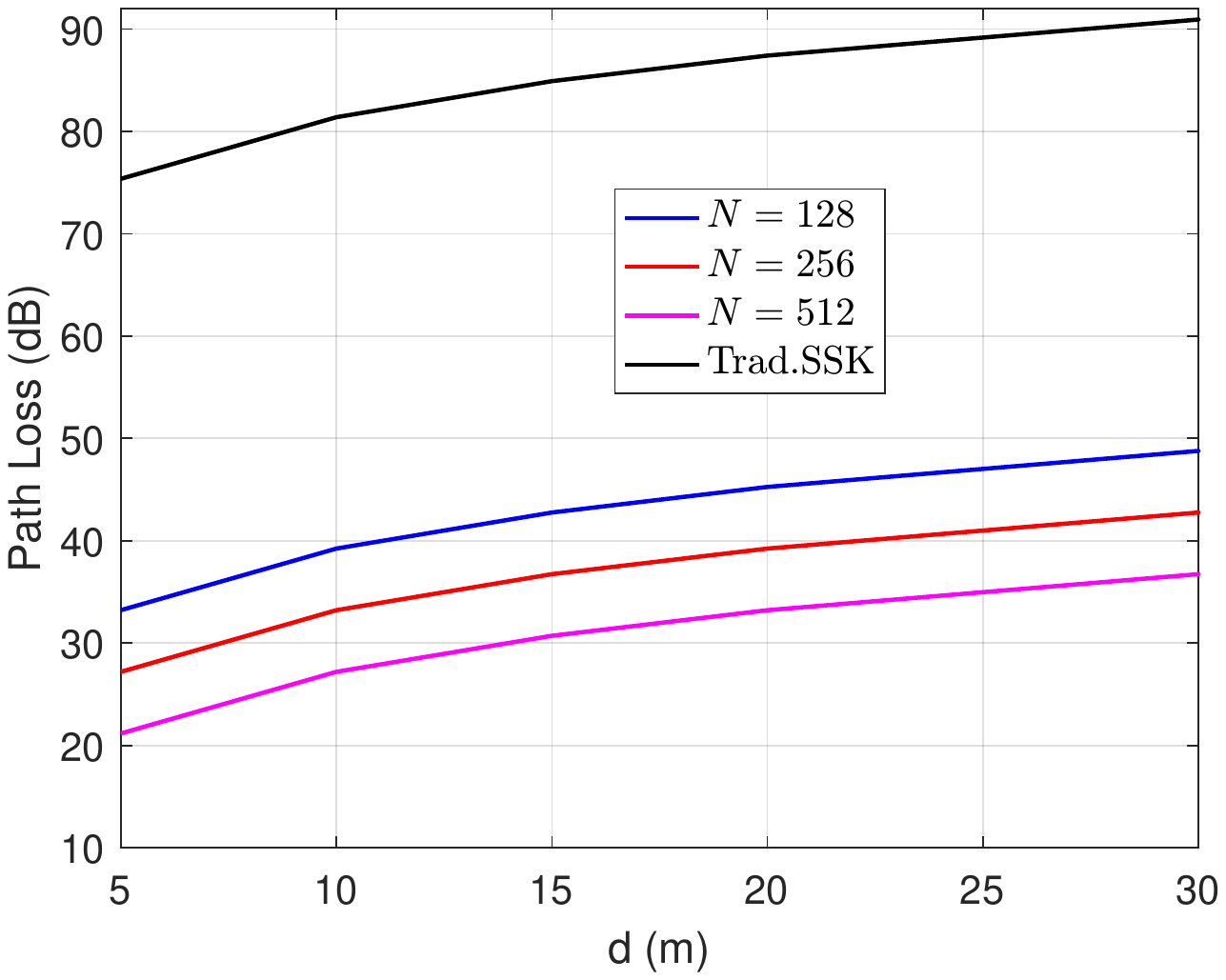} \hfill  \hspace{-5cm}
\includegraphics[height=11cm, width=7cm]{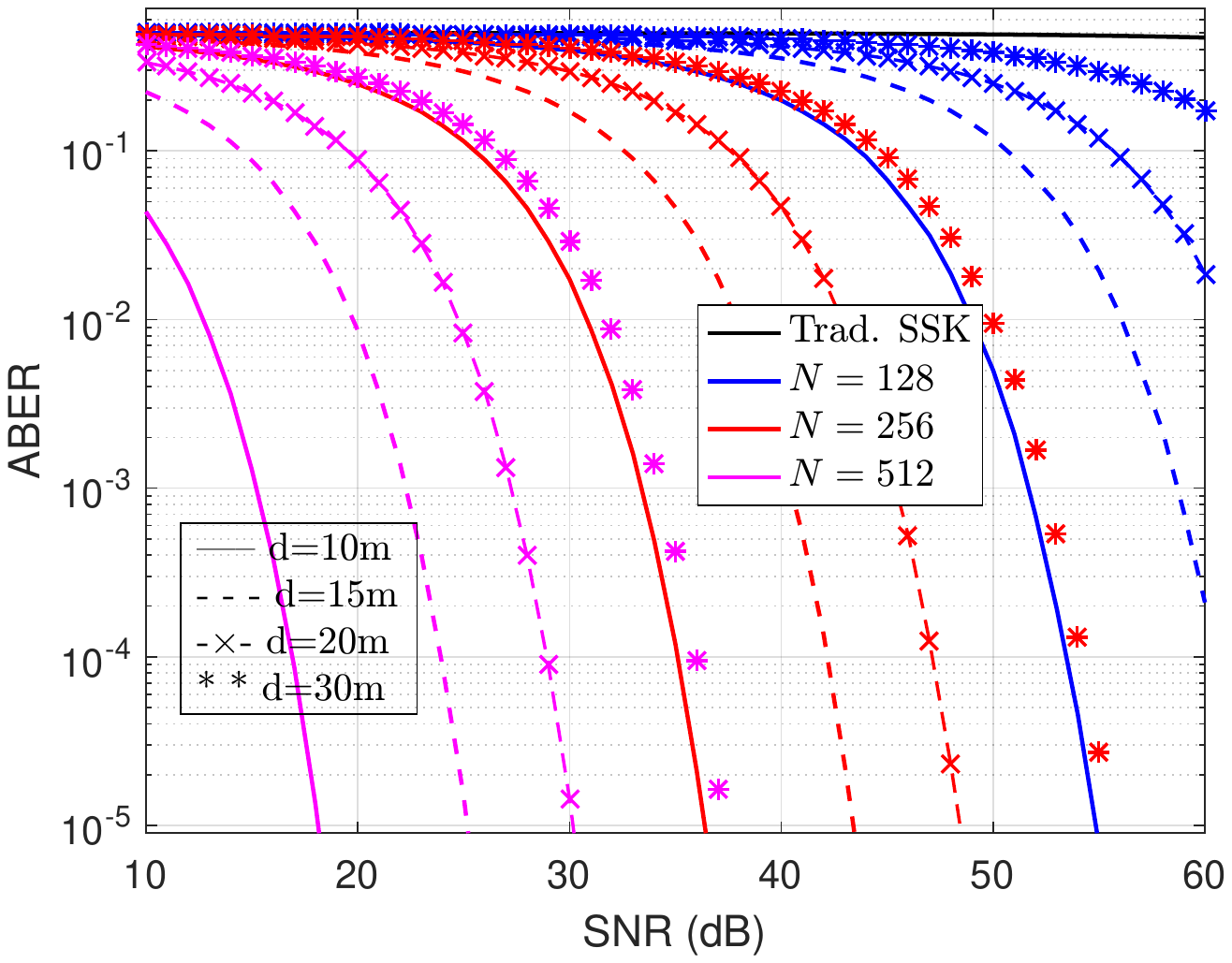} \hspace{-1.6cm} \vspace{-3.8cm}
\caption{Path loss effect on the intelligent RIS-SSK and the traditional SSK.} \vspace{-.4cm}
\end{figure}

\begin{figure}[t] \vspace{-4.5cm} \hspace{-.8cm}
\includegraphics[height=13.9cm, width=9.8cm]{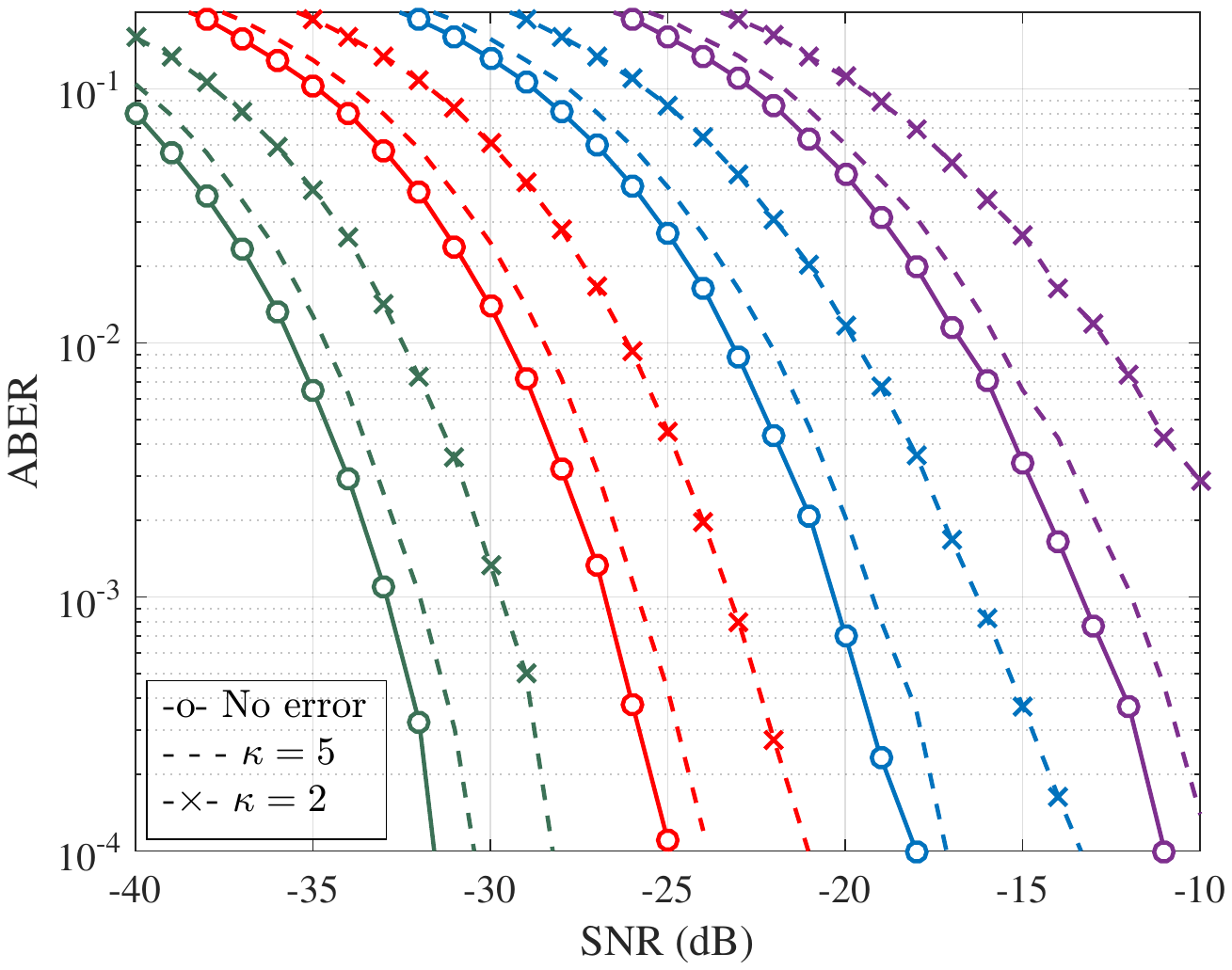} \vspace{-5.2cm}
\caption{ABER performance of the intelligent RIS-SSK scheme in the presence of channel phase estimation errors (purple: $N=32$, blue: $N=64$, red: $N=128$, green: $N=256$).}  \vspace{-.5cm} 
\end{figure}

%\begin{figure}[t] \hspace{-.3cm} \vspace{-.2cm}
%\includegraphics[height=6.2cm, width=8cm]{PathLossBER}
%\caption {Path loss effect on the proposed RIS-assisted SSK and traditional SSK for 28 GHz (left) and 73 GHz (right) frequency bands.} \vspace{-.3cm}
%\end{figure} 

\section{Conclusion}
A novel RIS-assisted IM concept, which provides ultra-reliable communications through the strategic adjustment of the RIS reflector phases, has been proposed in this letter as a potential, and rather promising, candidate for future wireless communication networks. Extensive computer simulations, confirmed by the analytical derivations, reveal that the RIS-based SSK scheme achieves very low error rates with unconventional energy efficiency, making it clearly superior to the reference designs. Additionally, the RIS-based SSK scheme may also avoid the need for complex coding methods to provide reliable communications, since it can be utilized effectively at extremely low SNR regions. Nevertheless, it is possible to achieve much better results by utilizing the advanced forms of the SSK in the system design. The development of new algorithms to make the RIS {\it intelligent} when it has partial channel state information, remains an interesting future research direction.  \vspace{-.3cm}

\bibliographystyle{IEEEtran}
\bibliography{ABC}

\end{document}